\documentclass[12pt]{article}
\usepackage{latexsym}
\usepackage{amsmath}
\usepackage{amssymb}
\usepackage{amsthm}
\usepackage{dsfont}
\usepackage{longtable}
\usepackage{color}
\usepackage{verbatim}
\usepackage[sort&compress,numbers]{natbib}
\usepackage{graphicx}
\usepackage{epstopdf}
\usepackage{float}
\usepackage{wrapfig}

\begin{document}

\markboth{A. M. Gavrilik and A. P. Rebesh} {Plethora of
$q$-oscillators possessing pairwise 
degeneracy}


\begin{center}
{\bf PLETHORA OF $q$-OSCILLATORS \\POSSESSING PAIRWISE ENERGY LEVEL DEGENERACY}

\vspace{10mm}

{\bf A. M. Gavrilik\footnote{omgavr@bitp.kiev.ua}, A. P. Rebesh}

\vspace{5mm}

{\it Bogolyubov Institute for Theoretical Physics, \\ 14-b Metrolohichna str., Kiev 03680, Ukraine}

\end{center}

\vspace{5mm}
\begin{abstract}
Using the  $q,\!p$-deformed oscillators as basic generating system,
we obtain diverse classes (which form distinct sectors of functional
continua) of novel versions of $q$-deformed oscillators, all of
which share the property of ``accidental'' degeneracy within a fixed
pair of energy levels $E_{m} = E_{m+1}$, $m\ge 1$, occurring at the
real deformation parameter fixed by an appropriate value $q(m)$ that
depends on $m$ and on particular model.
 Likewise, the degeneracy $E_{0} = E_{k} $ (where $k \geq 2$) takes
 place, for properly fixed $q=q(k)$, in most of those models.
The formerly studied model of $q$-oscillator known as the
Tamm-Dancoff cutoff deformed oscillator is contained in the continua
as isolated special case.

\vspace{3mm}
\noindent{\it Keywords:} $q,p$-deformed oscillators; (classes of)
nonstandard $q$-oscillators; pairwise energy levels degeneracy;
two-particle correlation intercept.

\vspace{2mm}
\noindent PACS Nos.: 02.20.Uw, 03.65.G, 03.65.Fd, 05.30.Pr

\end{abstract}

\section{Introduction}

For more then 15 years, diverse deformed models of quantum
oscillator play the important role in the study of modern quantum
mechanical systems, see e.g., Refs. \cite{Chang,Mizrahi} and
references therein. Diverse ($q$-,\ $q,\!p$-, etc.) deformed
oscillators, due to modified defining relations, can acquire
nontrivial and unusual properties essentially different from those
of the standard quantum oscillator. In Ref. 3, general
function-dependent deformations named the $f$-oscillators have been
studied including nonlinear coherent states and other physical
aspects.  Also, a unified framework for deformed single-mode
oscillators has been presented  in Ref. 4.
  Recently,  it has been                     demonstrated\cite{GR-1}
  that the two-parameter deformed $q,\!p$-oscillators
introduced in Ref.                             \cite{Chakr-Jag}
exhibit, at appropriate values of the deformation parameters $q$ and
$p$, the unusual property of so-called ``accidental'' double
(pairwise) degeneracy, within a fixed pair of energy levels. Such a
pair may be of the type $E_{m}=E_{m+1}$, of the type $E_{0}=E_{k}$,
\ $k\ge 2$, or even of more general type $E_{m_1}=E_{m_2}$, for
appropriate (completely determined) set of pairs of real values
$(q,p)$ such that both $q$ and $p$ depend on ${m_1},{m_2}$. Note
that this result extends to the two-parameter case the completely
analogous property (revealed earlier\cite{GR-2} and valid for
definite {\em real} values of the $q$-parameter) which is
characteristic for the rather special $q$-deformed oscillator, the
so-called 'Tamm-Dancoff (TD) cutoff' oscillator, which first
                appeared in           Refs.~\cite{Odaka,Jagan}.

It is important to emphasize that the just mentioned degeneracy
property, peculiar for the TD-oscillator, cannot occur in principle,
with real $q$, for the energy levels of the two most popular versions
of $q$-deformed oscillator, namely the
Biedenharn-Macfarlane\cite{Bied,Macf} (BM) $q$-oscillator and
Arik-Coon\cite{AC} (AC) version
of $q$-deformed oscillators\footnote{The situation is however
different, and some degeneracies can occur at {\em
phase-like} values of deformation parameter for
the BM  $q$-oscillator when $q$ is taken as a root of unity, see
e.g.,                                 Refs.  \cite{Bona,Wess} and
Ref. \cite{GR-1}.}.

Our goal in this paper is to demonstrate that, among {\it
one-parameter deformed} $q$-oscillator models, the TD $q$-oscillator
is not the unique and exotic one which possesses {\em at real
values} of deformation parameter $q$ the above mentioned property of
'accidental' double (or pairwise) degeneracy in a fixed pair of
energy levels. Just the contrary: we show there exist infinitely
many models of $q$-deformed oscillators (constituting in fact
functionally different continual sets or classes) all of which share
the degeneracy properties analogous to those mentioned above. The TD
$q$-oscillator is but a simplest individual case lying in one of
such classes.

Few words about the plan of the paper. In Sec. 2 we recapitulate
very shortly the basic setup concerning the $q,p$-deformed
oscillators, as well as the main facts about two-fold (pairwise)
degeneracy among their energy levels. In the next, 3rd section we
construct many classes of one-parameter deformed $q$-oscillators
which necessarily possess pairwise energy level degeneracy. In Sec.
4 we comment on a possible physical applications of the new
$q$-oscillators. Concluding remarks are contained in the last
section.

 \vspace{1mm}

\section{On the $q,p$-oscillators and their pairwise 'accidental'
degeneracies}

Our basic system is the two-parameter deformed $q,p$-oscillator
                                              algebra\cite{Chakr-Jag}
whose generating elements $A,$ $A^\dagger$ and ${\cal N}$
obey
\begin{equation}
 A A^\dagger - q\ A^\dagger A = p^{\cal N}  \ ,  \qquad\quad \quad
 A A^\dagger - p\ A^\dagger A = q^{\cal N} \ ,
\end{equation}
plus two more relations, involving $A$ (or $A^\dagger$) and ${\cal
N}$, similar to non-deformed case.

The pair of relations in (1), obviously symmetric under
$q\!\leftrightarrow\!p$, is satisfied with
\begin{equation}
A^\dagger A = [\![{\cal N}]\!]_{q,p} \ ,
\qquad  \qquad
A A^\dagger = [\![{\cal N}+1]\!]_{q,p} \ ,
                   \quad \quad
\end{equation}
  where the $q,p$-bracket means
\begin{equation}
[\![X]\!]_{q,p} \equiv \frac{ q^{X}-p^{X} }{ q-p } \
\end{equation}
for $X$ either an operator or a real number. Note that for a
non-negative integer $k$ the $q,\!p$-bracket in (3) reduces to the
(symmetric, homogeneous) $q,p$-polynomial:
\begin{equation}
[\![k]\!]_{q,p} = \frac{ q^{k}-p^{k} }{ q-p } =
\sum_{r=0}^{k-1} q^{k-1-r}p^r = q^{k-1}\sum_{r=0}^{k-1}q^{-r} p^r \
.
\end{equation}
At $p\!=\!1$ this two-parameter system reduces to the
                                                    AC-type\cite{AC}
$q$-oscillator, and putting $p=q^{-1}$ yields the other
distinguished                                      case\cite{Bied,Macf}
well known as the $q$-oscillator of Biedenharn and Macfarlane (or BM).
Finally, at $p=q$ the relations (1)-(4) turn into those of the TD
oscillator whose unusual properties were studied in Ref.
\cite{GR-2}.

For the Hamiltonian taken in the form
\begin{equation}
H  = \frac{1}{2} (A A^\dagger + A^\dagger A) \
\end{equation}
 the energy spectrum $H|n\rangle = E_n |n\rangle$ in the
$q,\!p$-deformed Fock space basis reads:
\begin{equation}
E_n =  \frac12 \Bigl( [\![n+1]\!]_{q,p}  +  [\![n]\!]_{q,p} \Bigr)
.
\end{equation}
With the account of (4), the energy spectrum takes the form
\begin{equation}
E_n =
\frac12 \left( q^{n} \sum_{l=0}^{n}q^{-l} p^l  + q^{n-1}
\sum_{s=0}^{n-1}q^{-s} p^s\right) .
\end{equation}
As $q,p\to 1$, we recover the familiar\
$E_n = n + \frac12$.  Moreover, $E_0 = \frac12$ for any $q,p$.

\vspace{1mm}
  We will consider the $q,p$-oscillators at real $q$, $p$
that belong to the intervals
\begin{equation}
0 \leq q\le 1\ ,    \qquad  \qquad  0\le p\le 1 \ .
\end{equation}
These define the quadrant in $q,p$-plane from which we exclude the
point (0,0).

\subsection{Degeneracy of the type $E_m=E_0$}
\vspace{1mm}

For the purposes of the present letter, of basic importance is the
possibility of 'accidental' degeneracies of $q,\!p$-oscillators,
obeying the relations (1)-(4).
  Let us first recapitulate the statements from         Ref. \cite{GR-1}
that will be needed in the sequel.

{\em Proposition 1.} There exists a continuum of pairs of values
($q, p$), or continuum of points of definite curve $F_{m\!,\,0}(p,q)=0$,
which provide the degeneracy
\begin{equation}
E_m - E_{0} = 0 \ , \hspace{14mm}  m\ge 2 \ .
\end{equation}

The degeneracy curve in $(q,\!p)$-plane, got from (9) and (7),
namely
\begin{equation}
F_{m\!,\,0}(q,p)\equiv\sum^{m}_{r=0}p^{m-r}q^{r} +
\sum^{m-1}_{s=0}p^{m-1-s}q^{s} - 1 = 0 \ ,
\end{equation}
implies certain implicit function $p=f_{m\!,\,0}(q)$ on the
$q$-interval in (8), continuous and monotonically decreasing.
   The latter can be                               shown\cite{GR-1}
through the derivative
\[
\frac{d p}{d q} = f'_{m\!,\,0}(q) = - \frac{\partial
F_{m\!,\,0}}{\partial q} \left(\frac{\partial F_{m\!,\,0}}{\partial
p}\right)^{-1} \ ,
\]
that is always negative. This continuous decreasing implicit function
$f_{m\!,\,0}(q)$ is represented by the curve (10), with endpoints on
the axes, in the quadrant (8).

\vspace{1mm}

 {\em Remark 1.}
The values of $q$ and $p$ from the pairs $(q,p)$ which solve Eq. (10),
in fact belong to the smaller, than in (8), intervals $0<q<q_m$ and
$0<p<p_m$ (here $p_m$ resp. $q_m$ solve (10) for $q=0$ resp. $p=0$,
and are such that $p_m,q_m<1$ and $p_m=q_m$).
   Moreover, denoting $q_\infty\equiv 1$ (since $q_m  \stackrel{m\to
\infty}{\longrightarrow} \ 1$) we have
\begin{equation}
q_2 < q_3 < q_4< ... < q_{m-1} < q_m < ... < q_\infty\ .
\end{equation}

Consider for a fixed $m$, \ $m\ge 2$,  the derivative
$f'_{m\!,\,0}(q)$ at the endpoints $(0,p_m)$,\ $(q_m,0)$ where
$q_m=p_m$, and at the midpoint (given by $p=q$) of each curve
$f_{m\!,\,0}(q)$.
  For any $m$ we have $f'_{m\!,\,0}(q)\vert_{q=p}=-1$ at the midpoint,
while at the endpoints the derivative depends on $m$, is negative,
and such that
\[
f'_{m\!,\,0}(q)\vert_{q=q_m\!,\, p=0} \ < \ -1 \ < \
f'_{m\!,\,0}(q)\vert_{q=0\!,\, p=p_m}\ < \ 0 \ .
\]

\vspace{1mm}
In Figure 1,  we illustrate the cases of $m=2,3,4,6$ by the curves
1, 2, 3, 4.

\vspace{1mm} \underline{Let $m=2$} (the case given by the curve 1 in
Fig.~1) The degeneracy implies
\[
F_{2\!,\,0}(q,p)=p^2+pq+q^2 +p+q-1 = 0 \ ,
\]
which yields the (explicit in this case) function
\begin{equation}
  p=f_{2\!,\,0}(q)=\frac{-1-q+\sqrt{(1+q)(1-3q)+4}}{2}
\end{equation}
monotonically decreasing for $0<q<q_2$ where $ q_2=(\sqrt5 -1)/2$,
and $p_2=q_2$.
Then,
\[
f'_{2\!,\,0}(q)= - \frac{p+2q+1}{2p+q+1}=
              \begin{cases}-\frac{p_2+1}{2p_2+1}\simeq-0.7236, & q=0 , \cr
                            \ \  - 1 ,           &  p=q , \cr
                             \frac{2q_2+1}{q_2+1}\simeq-1.382, & p=0 . \cr
              \end{cases}
\]


\subsection{Degeneracy of the type $E_{m+1}=E_m$}


Now consider the case with pairs of energy levels which are nearest
neighbors.

\vspace{1mm}
  {\em Proposition 2.} For each $m $, \  $m \ge 1$, there exists
a continuum of pairs of values $(q,p)$, or continuum of points of
definite curve $F_{m+1\!,\, m}(q,p)=0$ in the $(q,p)$-plane, for
which the degeneracy
\begin{equation}
\qquad E_{m+1} - E_m = 0\
\end{equation}
does hold\cite{GR-1}. The curve is given as
\begin{equation}
F_{m+1\!,\,m}(q,p)\equiv \sum^{m+1}_{r=0}p^{m+1-r}q^{r} -
\sum^{m-1}_{s=0}p^{m-1-s}q^{s} = 0 \ .
\end{equation}
Eq.~(14), due to polynomial form of its l.h.s., determines
a {\em continuous} implicit function $p\!=\!f_{m+1\!,\,m}(q)$.
  This function monotonically decreases
on the $q$-interval in (8) as follows from the inequality
\[
\frac{d p}{d q} = f'_{m+1\!,\, m}(q) =
- \frac{\frac{ \partial}{\partial q} F_{m+1\!,\, m}(q,p) } {
\frac{\partial}{ \partial p} F_{m+1\!,\, m}(q,p) }  < 0 \ .
 \]
Indeed, the both partial derivatives (being polynomials) are continuous and
positive functions of two variables.
E.g., in the case of \underline{$m=3$} due to eq.~(14) we have
\[
\hspace{1mm} F_{4,3}\equiv p^4+p^3q+p^2q^2+pq^3+q^4-p^2-pq-q^2=0
\]
from which we deduce that the inequality
\[
\frac{d p}{d q} = f'_{4\!,\, 3}(q) = -
   \frac{p^3+2qp^2+3q^2p+4q^3-p-2q}{q^3+2pq^2+3p^2q+4p^3-q-2p} < 0 \
   \]
is valid {\it at each point of the curve}. For more details
concerning the proof that $p\!=\!f_{m+1,m}(q)$ is continuous,
monotonically decreasing implicit function see Ref. \cite{GR-1}.

The cases  $m=1,2,4$ of $E_{m+1}-E_{m}=0$ are shown in Fig.~1 as the
curves 5,6,7.

\vspace{-2mm}

\begin{figure}  [hbtp]
\centering {
\includegraphics[angle=0, width=0.55\textwidth]   
{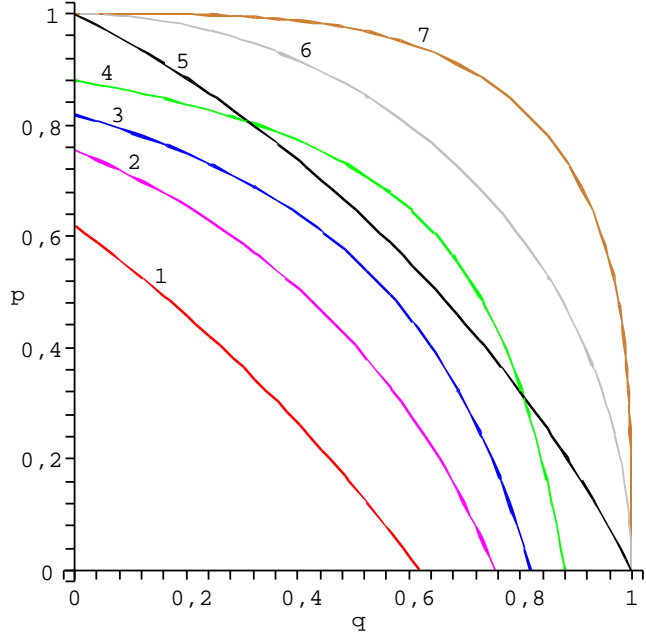} }
\caption{ Diverse cases of pairwise degeneracies of energy levels of
$q,\!p$-oscillator: the curves 1, 2, 3, 4 correspond to
$E_{0}=E_{2}$, $E_{0}=E_{3}$, $E_{0}=E_{4}$ and $E_{0}=E_{6}$, see
eq.~(9); the curves 5, 6, 7 correspond to $E_{1}=E_{2}$,
$E_{2}=E_{3}$ and $E_{4}=E_{5}$, see eq.~(13).}
\end{figure}

{\em Remark 2.} The case $m=1$ (or $E_1=E_2$) obviously differs from
the rest $m\ge 2$ cases since in the endpoints $(0,1)$ and $(1,0)$
the derivative $f'_{m+1\!,\, m}(q)$ acquires the values distinct
from those in  $m\geq 2$ cases. Indeed, at $m=1$, \
$f'_{2\!,\,1}(q)\vert_{q=0}=-\frac12$   and
 $f'_{2\!,\, 1}(q)\vert_{q=1}=-2$ \ (i.e., the derivative $f'_{2\!,\, 1}(q)$ continuously
drops from $-\frac12$ to $-2$ as $q$ runs from zero to one).
 In contrast, for each $m\ge 2$ we have $f'_{m+1\!,\, m}(q)\vert_{q=0}=0$ and
   $f'_{m+1\!,\, m}(q)|_{q\to 1} \to  - \infty$,
i.e., $f'_{m+1\!,\, m}(q)$ decreases from $0$ to $-\infty$ as $q$
grows from
 zero to one.     \
The distinction of $E_{1}=E_{2}$ case from the rest $m\ge 2$ cases,
e.g., $E_{2}=E_{3}$ or $E_{4}=E_{5}$, is evident in Fig.~1 (the
curve 5 versus the curves 6 and 7).

\vspace{1mm}
\section{ Novel $q$-oscillators with double degeneracies }

Thus, for $q,\!p$-deformed oscillators we have the property of
double degeneracy (within a fixed pair) of energy levels.
Due to this fact, a possibility  arises to obtain a host of
one-parameter $q$-deformed oscillators all inheriting, like the
TD-oscillator studied in                           Ref. \cite{GR-2},
the property of double (pairwise) degeneracy.
  Although there exist a plenty of continual classes of $q$-oscillators,
 below we present only few of them.

{\it $q$-Oscillators obtained by imposing $p=q^l$,
$0\!<\!l\!<\!\infty $.} The first large class of non-standard
$q$-oscillators can be obtained from the $q,\!p$-oscillators if we
impose $p=q^l$ with  $l$ such that $0\!<\!l\!<\!\infty $. Then we
have
\begin{equation}
A A^\dagger - q\ A^\dagger A = q^{l \cal N} \ ,   \qquad
 A A^\dagger - q^l \ A^\dagger A = q^{\cal N} \ ,
\end{equation}
\vspace{-6mm}
\[
A^\dagger A = [\![{\cal N}]\!]_{q,q^l}\ ,
                   \quad
                   \quad
                   \quad
A A^\dagger = [\![{\cal N}+1]\!]_{q,q^l}\ ,
\]
with the bracket as in (3), and the energy spectrum takes the form
\begin{equation}
E_n^{(l)} =
\frac12 \left( q^{n l}
+\bigl(1+q\bigr) q^{l(n-1)} \sum_{s=0}^{n-1} q^{s(1-l)}
\right) .  \ \ \
\end{equation}
 Clearly, this large class naturally splits in three subclasses:

(i) the one for which  $1 < l < \infty $;

(ii) the one for which  $0 < l < 1$;

(iii) the isolated case of $l=1$, equivalent to the TD-oscillator.

\noindent Any of these $q$-oscillators (see      Ref.~\cite{GR-2}
for the TD case) acquires the property of double degeneracy at certain
$q$, within a pair of energy levels.
This fact follows immediately since, in the ($q,p$)-plane, each line
$p=q^l$ with $l$ fixed so that $0\!<\!l\!<\!\infty$, crosses just
once the $q,\!p$-curve
$E_{m_i}-E_{m_j}=0$ of the corresponding degeneracy.


\begin{figure}  [hbtp]
\centering {
\includegraphics[angle=0, width=0.55\textwidth]   
{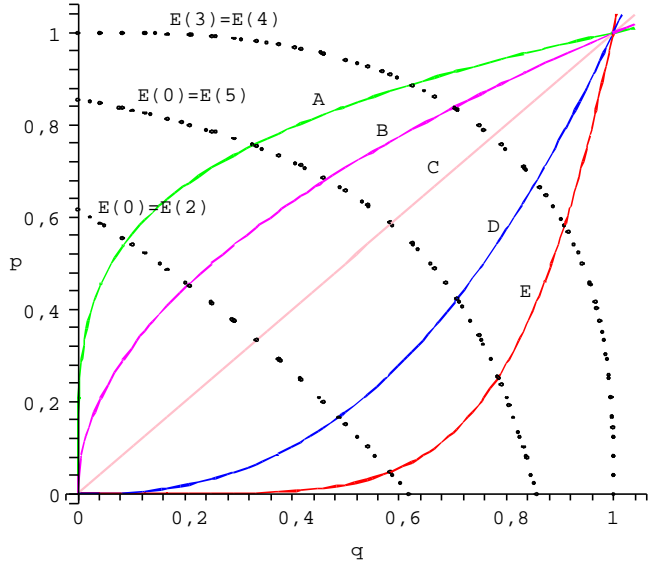} }
\caption{ Intersection of $E_0=E_2$,\ $E_0=E_5$,\ $E_3=E_4$
(degeneracy curves  of $q,\!p$-oscillator) by the lines $p=q^l$
leading to $q$-oscillators. Here $l=0.25,\ 0.5,\ 1,\ 2.5,\ 5.7$ for
A, B, C, D, E respectively.}
\end{figure}
\vspace{0.1mm}
 In Fig.~2, the representatives of all three subclasses are shown, including the $l=1$
(or the TD) case.
  Any representative of each subclass (i), (ii)
and (iii) does intersect exactly once with each of the degeneracy
curves $E_{m_i}-E_{m_j}=0$ at certain value of $q$.
  As a consequence
of the symmetry $q\leftrightarrow p$,  the first and the second
subclasses are 'dual' (or inverse) to each other.

{\it $q$-Oscillators deduced by setting $p=1+\alpha\ln q$, \
$\alpha>0$, and those deduced by setting
$p=\exp\,(\alpha(q\!-\!1))\,$,\ \ $\alpha>0$.} The next two classes
of $q$-deformed oscillators arise from $q,p$-oscillators if we
impose logarithmic or exponential type of relation $p=f(q)$.
 Let us first put
\begin{equation}
p=1+\alpha\ln q\, ,  \hspace{6mm} \alpha>0\, .
\end{equation}
 Then we arrive at the {\em logarithmic} class of
 $q$-oscillators, whose members
 are labeled by some real
$\alpha$,\ $\alpha>0$.
 Likewise, by imposing on the $q,\!p$-oscillators the relation
\begin{equation}
p=\exp\,(\alpha(q\!-\!1))\, ,  \hspace{6mm} \alpha>0\, ,
\end{equation}
 we get the {\em exponential} class of $q$-oscillators
 whose members are also labeled by $\alpha$. \
 Typical representatives of the two classes are shown in Fig.~3 and
Fig.~4.

\vspace{-3mm}
\begin{figure}  [hbtp]
\centering {
\includegraphics[angle=0, width=0.55\textwidth]   
{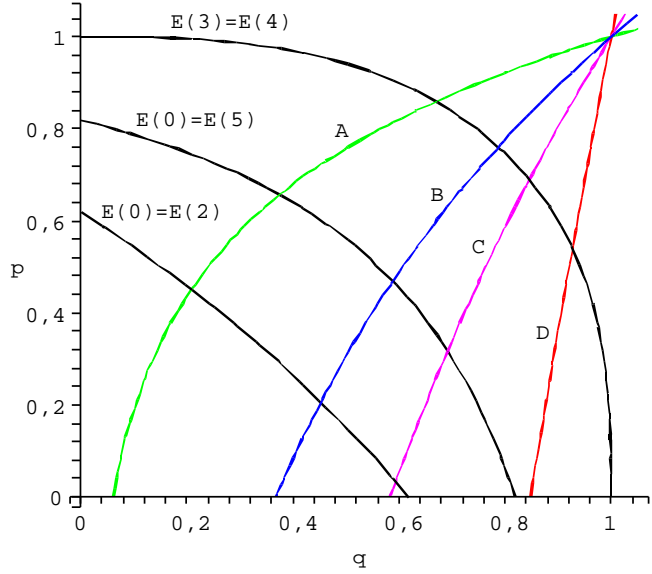} }
\caption{ Crossing
of $E_0=E_2$, $E_0=E_5$ and $E_3=E_4$ (degeneracy curves of
$q,\!p$-oscillator) by the lines $p=1\!+\!\alpha\ln q$ leading to
$q$-oscillators. Here A, B, C, D correspond to $\alpha=0.35,\ 1,\
1.85,\ 6.05$.}
\end{figure}

  {\em Remark 3.} As seen from Fig.~3, the
$q$-oscillator obtained from the 'master' $q,p$-oscillators with
$p=1+\alpha\ln q$ at $\alpha=6.05$, see line D, does not admit the
degeneracies $E_0=E_2$ and $E_0=E_5$ (and the 'intermediate' ones
$E_0=E_3$,  $E_0=E_4$): indeed, the line D does not cross those
degeneracy curves. Likewise, the $q$-oscillator obtained from the
'master' $q,p$-oscillators by imposing $p=\exp(\alpha(q-1))$ with
$\alpha=0.1653$, see the curve A in Fig.~4, does not admit the
degeneracies $E_0=E_2$ and $E_0=E_5$ (as well as $E_0=E_3$ and
$E_0=E_4$). This peculiarity is in sharp contrast with the class of
$q$-oscillators produced by $p=q^l$, see above, which admit all the
degeneracies occurring in the TD-oscillator and in
$q,p$-oscillators.

{\em Remark 4.} The expressions for energy $E_n=E(n)$ in the cases
of {\em exp}-class and {\em ln}-class of $q$-oscillators can be
explicitly given in complete analogy with Eq.~(16) of the $p=q^l$
class. Next, it is clear that due to the symmetry $q\leftrightarrow
p$ encoded in (3), (7), the roles of the deformation parameters $q$
and $p$ are interchangeable.
  Therefore, the {\em exp}-class and the {\em
ln}-class are dual (or inverse) to each other.

\vspace{-2mm}
\begin{figure}  [hbtp]
\centering {
\includegraphics[angle=0, width=0.55\textwidth]   
{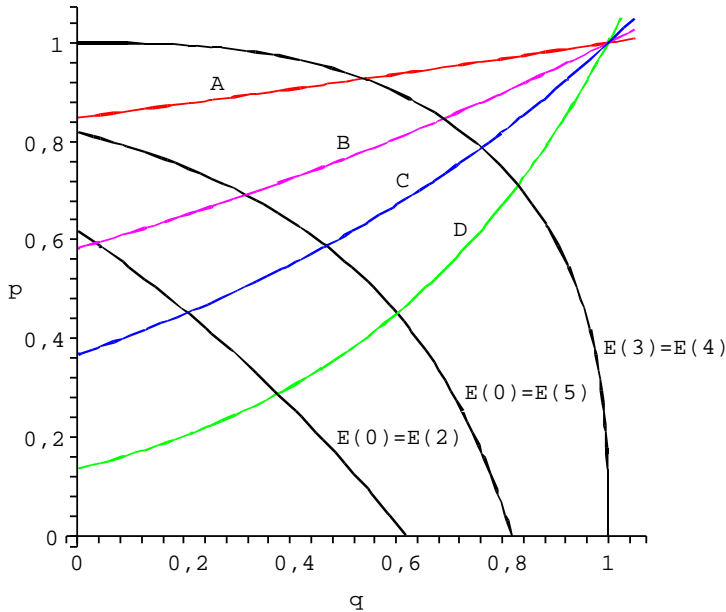} }
\caption{ Crossing
of $E_0=E_2$, $E_0=E_5$ and $E_3=E_4$ (degeneracy curves of
$q,\!p$-oscillator) by the lines $p\!=\!\exp(\alpha(q\!-\!1))$
leading to $q$-oscillators. Here A, B, C, D correspond to
$\alpha=0.165,\ 0.54,\ 1,\ 2$.}
\end{figure}

{\em Remark 5.}
Of course there exist, besides those given above,
  many other non-standard classes of one-parameter
  $q$-oscillators exhibiting analogous degeneracy properties.
To obtain any such class or a concrete desired nonstandard
$q$-oscillator by imposing an appropriate relation $p=f(q)$, we
should guarantee: ($i$) monotonic character of $p=f(q)$ in the
quadrant of $(q,p)$-plane given in (8);\, ($ii$) validity of the
property $f(1)=1$.\ Say, the relation $p=\frac{3}{q^2+1+q^{-2}}$
gives admissible example.

\section{Towards applications}

The first point to be emphasized is the behavior of  $n$-th level
 energy $E_n=E(n)$ as function of the particle number $n$.
   We find qualitative similarity between the shape of
$E(n)$ for the TD $q$-oscillator\cite{GR-2} and that of $E(n)$ for
the nonstandard versions of $q$-oscillator considered in the present
paper.
  Instead of trying to illustrate all possible cases, we
present in Fig.~5, for different $q$, the picture of $E(n)$ for one
typical member of the {\em exp}-class of $q$-oscillators, namely the
one got through setting $p=\exp (\frac12(q-1))$. As the Fig.~5
demonstrates, the basic features of $E(n)$ are the same as for the
TD deformed oscillator: (i) existence of maximum and hence the
ability to realize a degeneracy within certain pair of levels; (ii)
location of the maximum more and more to the right for successively
larger values of $q$ (compare the $q=0.4$ curve with those for
$q=0.7$ or $0.88$), with the tendency to turn into straight line
$E_n=n+\frac12$ (i.e., equal spacing of levels) of the standard
oscillator in the limit of $q=1$.

Let us emphasize once more the peculiar behavior, see Fig.~5, of
energy spectrum of the considered $q$-deformed oscillators versus
standard one. Namely, while the lowest value of energy is
$E_0=\frac{1}{2}$ for all the oscillators, both standard and
$q$-deformed, the energy $E_n$  of the considered $q$-oscillators,
after growing for few values $n\le n_0$ of the quantum number $n$,
becomes a decreasing function of $n$, with $E_n\rightarrow 0$ when
$n\rightarrow \infty$. This very special behavior of energy spectrum
can be of use in describing some exotic physical systems (maybe,
relativistic, and with complicated interaction).

\vspace{-3mm}
\begin{figure}  [hbtp]
\centering {
\includegraphics[angle=0, width=0.55\textwidth]   
{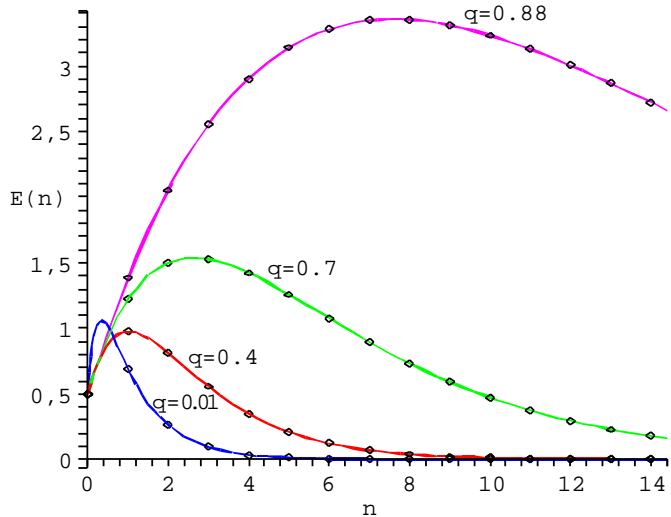} }
\caption{ The behavior of $E_n=E(n)$ of the $q$-oscillator obtained
by setting $p\!=\!\exp(\frac12(q\!-\!1))$. The four curves are given
for the values $q=0.01,\ 0.4,\ 0.7,\ 0.88$.}
\end{figure}

The other point is the application of new versions of
$q$-oscillators, or $q$-bosons, in the framework of respective
versions of $q$-Bose gas model and their use for description of
non-Bose features\cite{AGI-1,AGI-2,AGP,SIGMA} of multi-pion (and
multi-kaon) correlations observed in experiments on relativistic
heavy-ion collisions.
  For illustration, we give the picture confirming as well efficient
description using the presented classes of $q$-bosons. Namely, in
Fig.~6 we show the behavior of the intercept (maximal value)
$\lambda^{(2)}({\bf K})$ of two-pion momentum correlation function
derived\footnote{The results from the papers\cite{AdGa,SIGMA} are of
use in order to gain relevant formulas.} within the version of
$q$-Bose gas model based on the same version of $q$-oscillator
obtained through $p\!=\!\exp(\frac12(q\!-\!1))$.
 Three pairs of curves are shown, for the two temperatures 120 MeV
and 180 MeV in each pair. Like in Refs.~\cite{AGI-1,AGI-2}, the
intercept $\lambda^{(2)}({\bf K})$ rises with growing mean momenta
and its large momentum asymptotics shows saturation determined by
the parameter $q$.
  However, unlike the case of Arik-Coon $q$-bosons where
asymptotical value of intercept is just the $q$ by
itself\,\cite{AGI-1,SIGMA}, in the present version of $q$-Bose gas model
the intercept tends to its asymptotical value given as
$\lambda^{(2)}_{asym}= -1+q+\exp(\frac12(q\!-\!1))$.
Such feature, we hope, opens new possibilities for adjustment
to the appearing experimental data.

\vspace{-2mm}
\begin{figure}  [hbtp]
\centering {
\includegraphics[angle=0, width=0.55\textwidth]   
{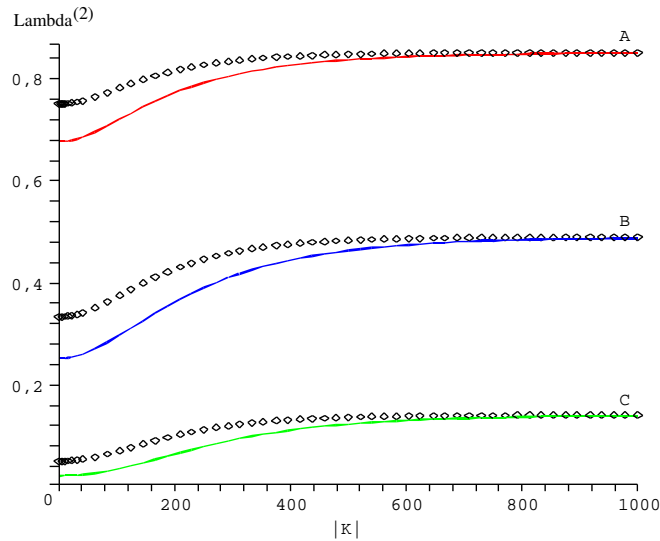} }
\caption{Intercept $\lambda^{(2)}({\bf K})$ of two-pion correlation
function in the $q$-Bose gas model that uses  
$q$-bosons based on setting $p\!=\!\exp(\frac12(q\!-\!1))$. The
pairs A, B, C of curves are given for 
$q=0.4,\ 0.65,\ 0.9$ respectively. Solid (dotted) curves correspond
to the temperature 180 MeV (120 MeV).}
\end{figure}

\vspace{-5mm}

\section{Concluding remarks}

 In                                          Ref.\,\cite{GR-2}
we have revealed the unusual ability of Tamm-Dancoff $q$-oscillator
to exhibit, at appropriate {\em real} values of $q$, the accidental
double degeneracy within a fixed pair of energy levels.
Subsequently, similar type of pairwise degeneracy of energy levels
has been demonstrated\cite{GR-1} in more general setting of
two-parameter $q,\!p$-oscillators if the admissible values of $q, p$
constitute certain continual set.

  It appears natural to exploit the $q,\!p$-oscillators with their
property of pairwise energy level degeneracies in order to obtain,
by definite reduction, many other unusual one-parameter deformed
oscillators with similar properties.\
 In this paper we presented a simple procedure that really allows to infer,
from the "master" $q,\!p$-oscillators, the innumerable sets (classes) of new
one-parameter models of $q$-deformed oscillators all of which
inherit, almost completely (cf. Remark 3), the same pairwise degeneracies,
e.g., the degeneracy $E_2=E_0$, or $E_2=E_1$, or $E_5=E_4$, or
others.

  Our results witness that the TD oscillator is not the unique one with
such exotic degeneracies, but only a separate case from a plethora of
continual classes of $q$-oscillators exhibiting the special property of the pairwise degeneracy
of energy levels within a fixed pair of them.
 As in the previous paper\cite{GR-2}, the degeneracy occurs at a
properly fixed value of the $q$-parameter, it concerns single pair
of levels and is an 'accidental' one (without underlying symmetry).

 Of course, there exist many other possibilities yielding
non-standard $q$-oscillators which can have even more complicated
patterns of degeneracy, e.g., those realized in Ref.~\cite{GR-3}
where the derived models exhibit two double (two pairwise)
degeneracies that involve two degenerate pairs: $E_{m_1}=E_{m_2}$
{\bf and} $E_{m_3}=E_{m_4}$. That means that the procedure developed
in this paper and in Ref.~\cite{GR-3} is really able to produce
myriads of one-parameter deformed nonlinear oscillators with more
and more nontrivial and unusual properties (patterns) of degeneracy.

We expect for the considered $q$-oscillators many interesting
physical applications.
 Say, it is of interest to explore in more detail the applicability
of such $q$-oscillators ($q$-bosons) in the context of effective
description of non-Bose properties of the two- and multi-pion
(-kaon) correlations observed in the experiments on relativistic
nuclear collisions. In Sec.~4, we have made a small step towards
possible applications. It was based on the version of $q$-bosons and
$q$-Bose gas involving one special member of the obtained {\em
exp}-class of nonstandard $q$-oscillators. Note in this context that
the earlier results on multi-particle correlations of $q,\!p$-bosons
in the two-parameter $q,\!p$-Bose gas model, including the explicit
formulas for (intercepts of) general $n$-particle momentum
correlation functions\cite{AdGa,SIGMA} are very useful. \
 Our conclusion is that the usage of the unusual $q$-oscillators, the main subject
of our present paper, in the framework of the corresponding $q$-Bose
gas model is, at least, on an equal footing with the other
distinguished versions of $q$-bosons (and $q$-Bose gas model), such as
Arik-Coon\cite{AC}, Tamm-Dancoff \cite{Odaka,Jagan} or
Biedenharn-Macfarlane\,\cite{Bied,Macf} ones.

\vspace{1mm}

{\bf Acknowledgements}

This work was partially supported by the Special Programme of
Division of Physics and Astronomy of NAS of Ukraine.

\end{document}